\documentstyle[prl,preprint,aps,epsf]{revtex}
\input{psfig}
\tightenlines
\def\bege{\begin{equation}}
\def\ende{\end{equation}}
\def\ddp{d\tilde{p}}
\def\ddl{d\tilde{l}}
\def\ddk{d\tilde{k}}
\def\gg{\gamma}
\def\mm{m_{\pi}}
\def\ra{\rangle}
\def\la{\langle}
\def\gt{\tilde{\gamma}}

\def\raw{\rightarrow}
\def\bdlra{\buildrel \leftrightarrow \over}
\def\bdra{\buildrel \rightarrow \over}
\draft
\begin{document}
\title{\bf Multiloop Diagrams in Non-Relativistic Field Theories and the Deuteron Quadrupole Moment}
\author{Michael Binger}
\address{ Department of Physics, North Carolina State University \\
          Raleigh, North Carolina 27695-8202 USA \\} \maketitle
\abstract{
A new method is developed to calculate multiloop Feynman diagrams in non-relativistic
 field theories. A consistent scheme for regularizing and renormalizing loop
integrals is established and shown to reproduce the results of dimensional regularization ($DR$)
and modified minimal subtraction ($\overline{MS}$) or DR and power divergence subtraction ($PDS$) 
up to next-to-leading order (NLO). However, significantly less effort is required to 
evaluate the integrals and the methods are easily generalized to higher order graphs.
Thus, even the most complicated multiloop graph can be expressed in terms of analytic functions. 
These techniques are then used to calculate the quadrupole moment 
of the deuteron to three loops in the Kaplan, Savage, and Wise effective field theory.
 A new unfixed direct $S{\raw}D$ wave counterterm occurs at this order
and its value is determined.} 

 {\newpage}

{\center{\bf{ I. INTRODUCTION}}}

Recent progress in effective field theory{\cite{1}} has made possible 
the calculation of many 
properties of nuclear systems in a perturbative expansion{\cite{2,3,4,5,6}}. The progress
is largely based on a consistent power counting developed by Kaplan, Savage, and Wise (KSW) 
which accounts for the large scattering lengths in the two nucleon sector by utilizing dimensional
regularization and a unique 
subtraction scheme called power divergence subtraction (PDS).
However, higher order 
graphs very quickly become intractable with the standard techniques of dimensional 
regularization and Feynman parameters. This is because combining three dimensional 
propagators in the usual way leads to square roots of polynomials of the 
Feynman parameters and one quickly becomes
entangled in a web of hopeless integrals.
While these may be computed numerically for finite
loop graphs, many of the more complicated divergent loop integrals 
remain elusive. Thus, one would like to find a new method for
 evaluating graphs
which avoids the shortcomings of these techniques and also allows for consistent regularization
 and renormalization schemes.  Furthermore, it would be beneficial to maintain the desirable KSW
  power counting. Other subtraction schemes {\cite{7,8}} have recently been developed that obey 
the KSW scaling, but none of these simplify the calculation of 
higher order contributions. 

  In this spirit, a 
 consistent regularization and subtraction scheme is developed 
for non-relativistic field theories. In particular, the KSW effective field theory
 is used throughout to develop the methods. 
The same recipe is applicable to all diagrams.
Loop integrals are transformed to coordinate space and divergences are transformed into
divergences in a parameter with units of energy, where they are consistently treated. 
The results are seen to be totally consistent with 
dimensional regularization and $PDS$ (or $\overline{MS}$) in the known cases. 
In addition, the coordinate space-parameter subtraction method, which we choose to call
$CPS$,  
is easily extended well
beyond the realm of the standard techniques in calculating higher order loop graphs. 
In section II the general methods are developed for bound state graphs.
We show how  calculate 2-pion exchange bound state graphs with an incident photon in the 
zero momentum limit, which is necessary for the NNLO calculation of the deuteron quadrupole moment.
 In section III, the results are generalized to finite photon
 momentum transfer graphs and 
above threshold scattering graphs. These examples serve to illustrate how
  graphs of any order may be obtained in terms of analytic functions.
 Section IV reviews the KSW effective field theory and the calculation of
 electromagnetic properties. Section V details the calculation of the deuteron
 quadrupole moment and the results are shown.  
 
   {\center{\bf{ II. BOUND STATE GRAPHS }}}

In Ref.[5], the deuteron electromagnetic form factors are calculated to NLO. Calculation of 
the full form factors at NNLO requires several graphs, the most difficult of which is the 2 pion
exchange with a finite momentum photon attached on one side (Fig.1). 
 First, we show how to evaluate the
  \begin{figure}
	\epsfxsize=3in
	\hbox to \hsize{\hss\epsffile{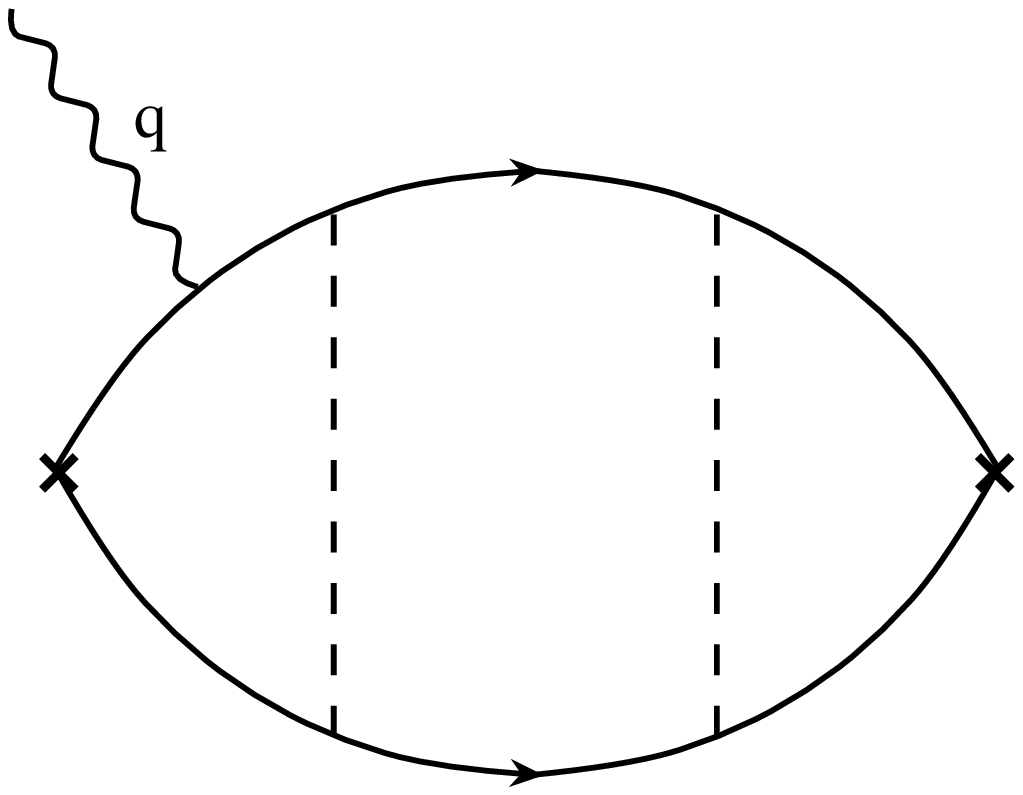}\hss}
	\label{fig:StrBrk}
 \end{figure}
 \begin{center}
 	{\small Fig.1 {\it Two potential pion exchange graph that gives a contribution to the deuteron quadrupole
	moment at NNLO. The photon corresponds to $A^0$.}}
 \end{center}
 graph at zero momentum
transfer. This will serve to introduce the $CPS$ method, since 
this complicated graph contains almost all possible classes of loop integrals that
 might occur in bound state problems.

 In the EFT of Kaplan, Savage, and Wise we find an amplitude
for Fig.1 of 
\bege
{\cal{A}}=\frac{-9g^4M_N^4e}{2f^4}{\int}\frac{d\tilde{p}d\tilde{l}d\tilde{k}
(2l_ik_jl{\cdot}k-l_il_jk^2-k_ik_jl^2+\frac{1}{2}l^2k^2{\delta}_{ij})}
{((p-q/2)^2+{\gamma}^2)(p^2+a^2)((p+l)^2+a^2)((p+l+k)^2+a^2)(l^2+m_{\pi}^2)(k^2+m_{\pi}^2)}
\ende
 where $d\tilde{p}=\frac{d^3p}{(2\pi)^3}$, $q$ is the photon 3-momenta, $p=\bdra{p}$, $l=\bdra{l}$, $k=\bdra{k}$,
 $\gamma=\sqrt{-EM_N}$, $m_{\pi}=138 {\rm MeV}$ is the pion mass, $a=\sqrt{{\gamma}^2+\frac{q^2}{4}}$,
  and $i$,$j$ are the initial and final deuteron spin indices, respectively.
We wish to extract the 
quadrupole moment part at zero momentum transfer. After straightforward
but tedious convolutions and projections, the numerator of eq.(1) can
  be written in terms of $p^2$, $l^2$, $k^2$, $(p-l)^2$, and $(p+k)^2$ as 
  54 independent terms such as  $(p-l)^4(p+k)^2l^2k^2/p^4$, $k^2l^4$, and other
  products of squares of the momentum variables.  Note the appearance of $1/p^4$ terms; these
  will introduce artificial infrared    
  divergences which must cancel in the end.
  After all these steps,
  one is left with integrals over products of 1, 3, 4, and 5 propagators. Note that the $(p-l)^2$ 
  propagator is raised to the fourth power from the Taylor expansion. 

    Let $I(n)$, $J(n)$, $K(n)$, and $L(n)$
    be 1, 3, 4, and 5 propagator integrals with the $(p-l)^2$ propagator raised to the
    $n$th power. We now explicitly show how to calculate these integrals in $CPS$.
   
      {\center {1-PROPAGATOR INTEGRALS}}
   
   By definition $I_{\gamma}(n){\equiv}{\int}d\tilde{p}\frac{1}{(p^2+{\gamma}^2)^n}$. Transforming to 
   coordinate space we have 
   \bege
   I_{\gamma}(1)={\int}d\tilde{p}{\int}d^3xG_{\gamma}(x)e^{-i\bdra{p}{\cdot}\bdra{x}},
   \ende
    where $G_{\gamma}(x)=\frac{e^{-{\gamma}x}}{4{\pi}x}$ is the Fourier transform of the propagator.
    Trivially $I_{\gamma}(1)=G_{\gamma}(0)=\frac{1}{4\pi}(1/x-\gamma)|_{x=0}$.
    Labelling the divergence $\mu$ we have 
    \bege
    I_{\gamma}(1)=\frac{\mu-\gamma}{4\pi}.
    \ende
    Alternatively we could have differentiated with respect to $\gamma$ and integrated back,
    yielding $I_{\gamma}(1)=-{\int}_{\infty}^{\gamma}d{\gamma}2{\gamma}{\int}d^3x(G_{\gamma}(x))^2$,
    which becomes $I_{\gamma}(1)=-{\int}_{\infty}^{\gamma}d{\gamma}\frac{1}{4\pi}=
    \frac{{\gamma}|_{\infty}-\gamma}{4\pi}$. Labelling the linear parameter divergence
    as $\mu$ we have eq.(3). In fact, eq.(3) is exactly what would be obtained using 
    dimensional regularization and $PDS$. This is a general result : poles in 3 dimensions
    using dimensional regularization and $PDS$ are linear parameter divergences in $CPS$. Other $I_{\gamma}(n)$ for
    $n>1$ are 
    obtained easily by differentiating eq.(3) with respect to $\gamma$. We also have integrals of
     the form
    $IO{\equiv}{\int}d\tilde{p}\frac{1}{p^2}$, for which we let ${\gamma}{\rightarrow}0$ in 
    the above. Hence,
    \bege
    IO=\frac{\mu}{4\pi}.
    \ende
     The standard $DR$ dictum that scaleless integrals are zero seems to 
    indicate that $CPS$ is not equivalent to dimensional regularization and $PDS$ in eq.(4). However,
     scaleless integrals should only be zero when 4 dimensional logarithmic divergences
    are considered, such as in $\overline{MS}$, since then there is only one scale
    (the renormalization scale) with which to make a dimensionless argument for the 
    logarithm. But for power law divergences, as in PDS, all we need is one 
    scale, the renormalization scale. Thus, the natural extension of dimensional regularization to PDS 
    is equivalent to eq.(4). Finally, we consider $IZ{\equiv}{\int}d\tilde{p}\frac{1}{p^4}$.
    After Fourier transforming and performing the integrals, we are left with 
    $IZ=\frac{1}{4\pi}{\int}_0^{\infty}dx$. Labelling the linear coordinate divergence
    as $\nu$ we have 
    \bege
    IZ=\frac{\nu}{4\pi}.
    \ende
    The same result is obtained by letting ${\gamma}{\rightarrow}0$ in $I(2)=\frac{1}
    {8{\pi}{\gamma}}$ and then $\frac{1}{\gamma}|_{{\gamma}=0}{\rightarrow}2\nu$. 
    For our purposes it is not really important how one labels these artificial infrared divergences since a
    conspiracy amongst the 54 terms must cancel them in the end.
    {\center{3-PROPAGATOR INTEGRALS}}
    
    By definition 
    \bege
    J(n){\equiv}{\int}d\tilde{p}d\tilde{l}
    \frac{1}
    	{ (p^2+{\gamma}^2) ((p-l)^2+{\gamma}^2)^n (l^2+m_{\pi}^2) }. 
    \ende
    In coordinate space we have 
    \bege
    J(1)={\int}{\ddp}{\ddl}d^3xd^3yd^3zG_{\gamma}(x)G_{\gamma}(y)G_{m_{\pi}}(z)
    		e^{-i[{\bdra p}{\cdot}{\bdra x}+({\bdra p}-{\bdra l}){\cdot}{\bdra y}+{\bdra l}{\cdot}{\bdra z}]}.
    \ende
    This quickly reduces to $J(1)= 
    {\int}d^3xd^3yd^3zG_{\gamma}(x)G_{\gamma}(y)G_{m_{\pi}}(z){\delta}(\bdra{x}+\bdra{y}){\delta}(\bdra{y}-\bdra{z})
    =\frac{1}{16{\pi}^2}{\int}_0^{\infty}dx\frac{e^{-bx}}{x}$, 
    where $b=2\gamma+m_{\pi}$. Differentiating with respect to $b$, we have
    $J(1)=\frac{-1}{16{\pi}^2}{\int}_{\infty}^{b}db\frac{1}{b}=\frac{-1}{16{\pi}^2}[\log
    {b}-\log{b|_{\infty}}]$. Hence,
    \bege
    J(1)= \frac{-1}{16{\pi}^2}\log{\frac{2\gamma+m_{\pi}}{\mu}}.
    \ende
     Again, this is the same result obtained using $DR$
     and $PDS$ or $\overline{MS}$, but those require substantially more effort.
     In general, four dimensional logarithmic divergences of dimensional regularization 
     become logarithmic parameter divergences in $CPS$.
     Using the same methods we can derive 
     \begin{eqnarray}
     JO(1)&{\equiv}& {\int}d\tilde{p}d\tilde{l}\frac{1}{p^2((p-l)^2+{\gamma}^2)(l^2+m_{\pi}^2)}
     \nonumber\\
     &=&\frac{-1} {16{\pi}^2} \log {\frac {{\gamma}+m_{\pi}} {\mu} }.
    \end{eqnarray}
    and 
    \begin{eqnarray}
    JZ(1)&{\equiv}& {\int}d\tilde{p}d\tilde{l}\frac{1}{p^4((p-l)^2+{\gamma}^2)(l^2+m_{\pi}^2)}
    \nonumber\\
    &=&\frac{2{\nu}(m_{\pi}+\gamma)-1}{32{\pi}^2(\gamma+m_{\pi})^2},
    \end{eqnarray}
    where $\nu$ is defined after eq.(5).
    Of course we get $J(n)$, $JO(n)$, and $JZ(n)$ for $n>1$ by differentiation with
    respect to $\gamma$.

     {\center{4-PROPAGATOR INTEGRALS}}

     There are several different classes of 4-propagator integrals. The first is 
     \bege
     K(n){\equiv} {\int}{\ddp}{\ddl}{\ddk}\frac{1}{((p-l)^2+{\gamma}^2)^n((p+k)^2+{\gamma}^2)(l^2+m_{\pi}^2)(k^2+m_{\pi}^2)},
     \ende
     which contains both 3 and 4 dimensional poles in dimensional regularization.
      In configuration space we obtain
     $K(1)= {\int}d^3x(G_{\gamma}(x))^2(G_{m_{\pi}}(x))^2
          = \frac{1}{64{\pi}^3}{\int}_0^{\infty}dx\frac{e^{-{\alpha}x}}{x^2}$, where $\alpha=2(m_{\pi}+\gamma)$.
	  By partial integration
	  $K(1)=\frac{1}{64{\pi}^3}[-\frac{e^{-{\alpha}x}}{x}|_0^{\infty}
	  -{\alpha}{\int}_0^{\infty}dx\frac{e^{-{\alpha}x}}{x}]$. Hence the divergences
	  are manifested as the linear and logarithmic poles seen before
	   in eq.(3) and eq.(8).
      The result is
      \bege
      K(1)=\frac{1}{64{\pi}^3}[2(m_{\pi}+\gamma)(\log{\frac{2(m_{\pi}+\gamma)}{\mu}}-1)+\mu].
      \ende     
      Next we consider integrals of the form
      \bege
      KP(n) {\equiv} {\int}{\ddp}{\ddl}{\ddk}\frac{p^2}{((p-l)^2+{\gamma}^2)^n((p+k)^2+{\gamma}^2)(l^2+m_{\pi}^2)(k^2+m_{\pi}^2)}.
     \ende
   This is greatly complicated by the appearance of $p^2$ in the numerator which cannot be decoupled
   with momenta in the denominator. But one can evaluate the integral simply by 
   using $p^2={\int}d^3x{\delta}(\bdra{x})(-{\nabla}^2_x)e^{-i{\bdra p}{\cdot}{\bdra x}}$. It follows that
   \bege
   KP(1)=-{\int}d^3x d^3y d^3z G_{\gamma}(y)G_{\gamma}(z)G_{m_{\pi}}(y)G_{m_{\pi}}(z){\delta}
   	(\bdra{x}){\nabla}^2_x{\delta}(\bdra{x}+\bdra{y}+\bdra{z})
   \ende
   where the derivative operator acts only on the delta function.
   Now we use ${\int}d^3xf(\bdra{x}){\nabla}^2_x{\delta}(\bdra{x})=f''|_{x=0}$, where the primes denote 
   operation by ${\nabla}^2_x$. Thus,
   \begin{eqnarray}
   KP(1) &=& -{\int}d^3yd^3z G_{\gamma}(y)G_{\gamma}(z)G_{m_{\pi}}(y)G_{m_{\pi}}(z){\delta}''(\bdra{y}+\bdra{z})
   	\nonumber\\
       	 &=& -{\int}d^3y G_{\gamma}(y)G_{m_{\pi}}(y)[G_{\gamma}(y)G_{m_{\pi}}(y)]''
   \end{eqnarray}
   From the definition of $G_{\gamma}(x)$ we find 
   \begin{eqnarray}
    \bdra{\nabla}_xG_{\gamma}(x)&=&0
      \nonumber\\
    {\nabla}^2_xG_{\gamma}(x)&=&{\gamma}^2G_{\gamma}(x)-{\delta}(\bdra{x}).
   \end{eqnarray}
   Straightforwardly,
   \begin{eqnarray}
   KP(1) &=& G_{\gamma}(y)(G_{m_{\pi}}(y))^2|_{y=0}+(G_{\gamma}(y))^2G_{m_{\pi}}(y)|_{y=0}
         -({\gamma}^2+m_{\pi}^2){\int}d^3y(G_{\gamma}(y))^2(G_{m_{\pi}}(y))^2
  \nonumber\\
  &=& I_{\gamma}^2(1)I_{m_{\pi}}(1)+I_{\gamma}(1)I_{m_{\pi}}^2(1)-({\gamma}^2+m_{\pi}^2)K(1).
  \end{eqnarray}
   We must also deal with $KPP(n)$, which is the same as $KP(n)$ except the numerator is $p^4$ instead of
   $p^2$. This is obtained in similar fashion and the only new result needed is ${\nabla}^4_xG_{\gamma}(x)=
   {\gamma}^4G_{\gamma}(x)-{\gamma}^2{\delta}(\bdra{x})-{\delta}''(\bdra{x})$. We obtain 
   \bege
   KPP(1)=({\gamma}^2+m_{\pi}^2)^2K(1)-2(2{\gamma}^2+m_{\pi}^2)I_{\gamma}^2(1)I_{m_{\pi}}(1)-2({\gamma}^2+2m_{\pi}^2)I_{\gamma}(1)I_{m_{\pi}}^2(1).
   \ende
   
      {\center{5-PROPAGATOR INTEGRALS}}

   Finally, we come to the master integrals for 2 pion exchange graphs. By definition
   \bege
   L(n){\equiv}{\int}{\ddp}{\ddl}{\ddk}\frac{1}{(p^2+a^2)((p-l)^2+{\gamma}^2)^n((p+k)^2+{\gamma}^2)(l^2+m_{\pi}^2)(k^2+m_{\pi}^2)},
   \ende
   where $a$ will be set to $\gamma$ in the end, but we label it differently for reasons that will be clear.
   After the delta function integrations in coordinate space we arrive at
   \begin{eqnarray}
   L(1)&=& \frac{1}{(4{\pi})^5}{\int}d^3xd^3y\frac{e^{-(\gamma+m_{\pi})(x+y)-a|\bdra{x}-\bdra{y}|}}{x^2y^2|\bdra{x}-\bdra{y}|}
      \nonumber\\
       &=& \frac{1}{2a(4{\pi})^3}{\int}_0^{\infty}dx{\int}_0^{\infty}dy 
           \frac{e^{-(\gamma+m_{\pi})(x+y)}}{xy}[e^{-a|x-y|}-e^{-a(x+y)}]
       \nonumber\\
       &{\equiv}&  \frac{1}{2a(4{\pi})^3}I_L,	    
   \end{eqnarray}
    where we have performed the angular integrations in the last step and defined $I_L$ in the obvious manner.
    Now use the symmetry between $x$ and $y$ to write
    \begin{eqnarray} 
    I_L &=& 2{\int}_0^{\infty}\frac{dx}{x}{\int}_0^x\frac{dy}{y}e^{-(\gamma+m)(x+y)}[e^{-a(x-y)}-e^{-a(x+y)}] \nonumber\\
    	&{\equiv}& I_L^A-I_L^B.
    \end{eqnarray}
    One must be careful here. We used the property that the integrand is symmetric in $x{\leftrightarrow}y$ to get rid of 
    the troublesome absolute value. But this must be done for both pieces, since $L(1)$ is finite but
    each individual piece $I_L^A$ and $I_L^B$ is divergent. The wrong answer would be obtained by exploiting the 
    symmetry in $I_L^A$, but not $I_L^B$. This is a consequence of the following simple rule : only finite integrals may be 
    exploited for their symmetry properties as above, since otherwise information is lost about the relationship between
     the artificial divergences.  In splitting a finite integral into several divergent pieces, every piece must be 
     dealt with in the same manner.
    Now $I_L^A=2{\int}_0^{\infty}\frac{dx}{x}{\int}_0^x\frac{dy}{y}e^{-bx-\tilde{m}y}$, where $b=m_{\pi}+\gamma+a$ and
     $ \tilde{m}=m_{\pi}+{\gamma}-a $. We only need evaluate $I_L^A$ explicitly since  
    $I_L^B=I_L^A(\tilde{m}{\rightarrow}b)$. Differentiating with respect to $b$ and $\tilde{m}$ we have 
    \begin{eqnarray}
    I_L^A &=& -2{\int}_{\infty}^{\tilde{m}}d\tilde{m} {\int}_0^{\infty}\frac{dx}{x} {\int}_0^xdy e^{-bx-{\tilde{m}}y} \nonumber\\
    	  &=& -2{\int}_{\infty}^{\tilde{m}} \frac{d{\tilde{m}}} {\tilde{m}} {\int}^b_{\infty}db {\int}_0^{\infty}dx e^{-bx}(e^{-{\tilde{m}}x}-1) \nonumber\\     
	  &=& -2{\int}_{\infty}^{\tilde{m}} \frac{d{\tilde{m}}} {\tilde{m}} {\int}^b_{\infty}db (\frac{1}{{\tilde{m}}+b}-\frac{1}{b}) \nonumber\\
	  &=& -2{\int}_{\infty}^{\tilde{m}} \frac{d{\tilde{m}}} {\tilde{m}} (\log {\frac{\tilde{m}+b}{\mu}} - \log{\frac{b}{\mu}}),
    \end{eqnarray}
    where we have used the $CPS$ prescription for the artificial logarithmic divergences. Next we employ 
    \bege
    {\int}dx\frac{\log{(x+y)}}{x}=\log{x}\log{y}-{\rm Polylog}_2({-\frac{x}{y}}),
    \ende
    whence we deduce
    \bege
    I_L^A=2{\rm Polylog}_2({-\frac{{\tilde{m}}}{b}}){\Bigl{|}}^{\tilde{m}}_{\infty}.
    \ende
     The polylogarithms are generalized logarithms and may be defined by 
     ${\rm Polylog}_n(x)=\sum^{\infty}_{i=1}\frac{x^i}{i^n}$. 
     It can easily be shown that for large $x$,
      ${\rm Polylog}_2(-x){\rightarrow}-\frac{(\log{x})^2}{2}-\frac{{\pi}^2}{6}$.
     Thus, 
     \bege
     I_L^A=2[{\rm Polylog}_2({\frac{-{\tilde{m}}}{b}})+\frac{(\log{\frac{b}{\mu}})^2}{2}+\frac{{\pi}^2}{6}]
     \ende
     and from this we obtain $I_L^B=2[\frac{(\log{\frac{b}{\mu}})^2}{2}+\frac{{\pi}^2}{12}]$, so that
     $I_L=2[{\rm Polylog}_2({\frac{-{\tilde{m}}}{b}}) + \frac{{\pi}^2}{12}]$. Hence, we finally have
     \begin{eqnarray}
     L(1)&=& \frac{1}{(4{\pi})^3a} [{\rm Polylog}_2({\frac{a-m_{\pi}-\gamma}{a+m_{\pi}+\gamma}})+\frac{{\pi}^2}{12}]
     \nonumber\\
     	 &=& \frac{1}{(4{\pi})^3{\gamma}} [{\rm Polylog}_2({\frac{-m_{\pi}}{m_{\pi}+2\gamma}})+\frac{{\pi}^2}{12}]
     \end{eqnarray}
     $L(2)$ may be obtained by differentiating this with respect to $\gamma$. One might hope to get $L(n)$ for $n>2$
     similarly, but because the $x{\leftrightarrow}y$
     symmetry was required, we can not label the 2 $\gamma$'s in eq.(19) differently and then differentiate. However, we can obtain 
     $L(3)$ and $L(4)$ by differentiating eq.(26) with respect to $\gamma$ and using the following integral which is easily calculated :
     \begin{eqnarray}
      L22&{\equiv}&{\int}{\ddp}{\ddl}{\ddk}\frac{1}{(p^2+a^2)((p-l)^2+{\gamma}^2)^2((p+k)^2+{\gamma}^2)^2(l^2+m_{\pi}^2)(k^2+m_{\pi}^2)}
       \nonumber\\
      &=& \frac{1}{8{\gamma}^2(4{\pi})^3(m_{\pi}+a+\gamma)^2(m_{\pi}+\gamma)}.
      \end{eqnarray}
      By explicit calculation or by taking $a{\rightarrow}0$ in eq.(26) we arrive at
      \begin{eqnarray}
        LO(1) &{\equiv}& {\int}{\ddp}{\ddl}{\ddk}\frac{1}{p^2((p-l)^2+{\gamma}^2)((p+k)^2+{\gamma}^2)(l^2+m_{\pi}^2)(k^2+m_{\pi}^2)}
	   \nonumber\\
	      &=& \frac{2\log{2}}{(4{\pi})^3(m_{\pi}+\gamma)}.
      \end{eqnarray}
     Similarly, recall $\frac{1}{a}|_{a=0}{\rightarrow}2{\nu}$, and find 
     \begin{eqnarray}
      LZ(1) &{\equiv}& {\int}{\ddp}{\ddl}{\ddk}\frac{1}{p^4((p-l)^2+{\gamma}^2)((p+k)^2+{\gamma}^2)(l^2+m_{\pi}^2)(k^2+m_{\pi}^2)}
	   \nonumber\\
	    &=& \frac{1}{(4{\pi})^3(m_{\pi}+\gamma)^3}[{\nu}(m_{\pi}+\gamma)-\frac{2\log{2}+1}{3}].
      \end{eqnarray}	    
     $LO(n)$ and $LZ(n)$ for $n>1$ are obtained similarly. Note that $L(n)$ and $LO(n)$ are finite and 
     $LZ(n)$ contains artificial infrared divergences, as expected. 

      This exhausts our treatment of the integrals necessary to calculate Fig.1. A few comments are in order. 
     In the $CPS$ system the regularization and subtraction are done in the same step. The scale $\mu$
     can be viewed as a label for the divergence and also as the renormalization scale.
     The counterterms introduced to subtract the infinities satisfy the local symmetries of 
     the Lagrange density. Specifically, gauge invariance and Euclidean invariance are easily seen to be
     preserved by the CPS regulator because the poles in CPS are isomorphic with poles in
      dimensional regularization, at least to the order we are working. 
      This equivalence also implies that the KSW power counting scaling is preserved and the CPS
      scheme may be consistently implemented with other PDS calculations. Particularly,
      in section V the previously determined numerical values for the NLO counterterm 
      coefficients $C_2$ and $D_2$ {\cite{1}} are used with CPS calculated amplitudes for the deuteron 
      quadrupole moment.  
       Also note that 
     the derivation of these integrals does not change as we go above threshhold, so analogous results
      can be obtained with the prescription ${\gamma}{\rightarrow}-ip$.
     	 The application of these 
     methods to higher order diagrams, such as the 3 potential pion exchange graph analogous to Fig.1, is 
     straightforward. For example, the master integral for the 3 pion exchange graph, involving
     seven propagators, is expressed in terms of ${\rm Polylog}$ of order 3 and lower. In general,
     an $n$-pion exchange graph can be expressed in terms of ${\rm Polylog}_n$ and lower order 
     polylogarithms.  
     
     \begin{figure}
	\epsfxsize=3in
	\hbox to \hsize{\hss\epsffile{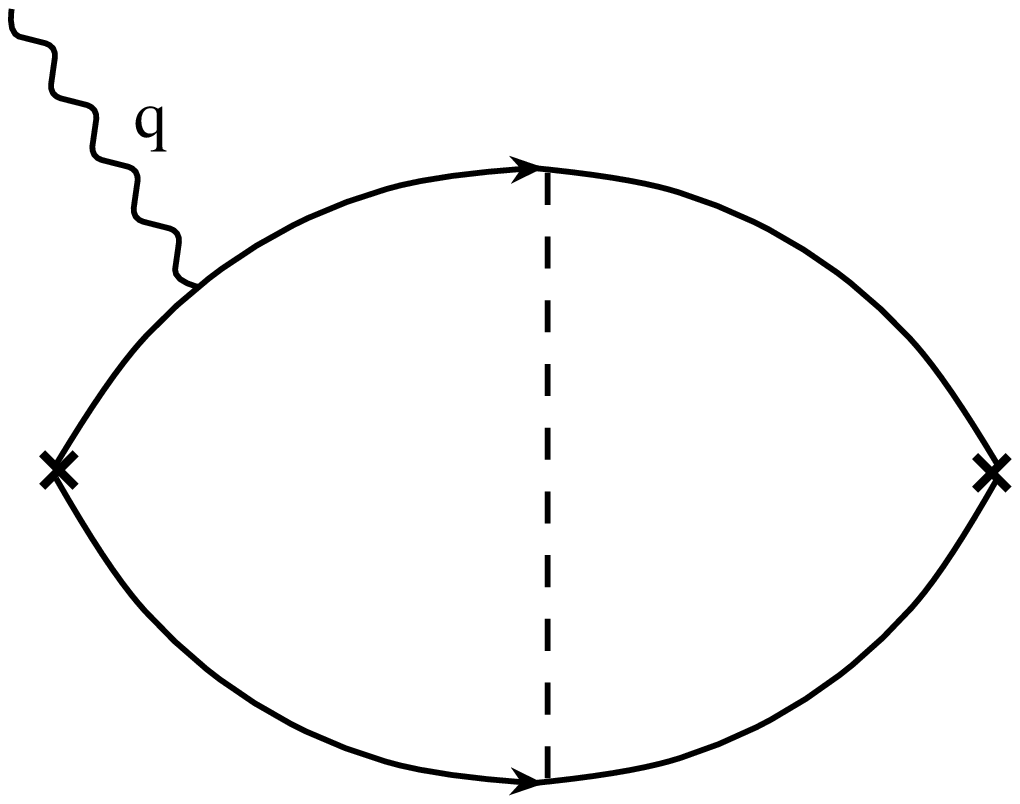}\hss}
	\label{fig:StrBrk}
	\end{figure}
	 \begin{center}
 	{\small Fig.2.{\it  Photon coupled to a deuteron bound state with one pion exchange. This gives
	    the deuteron quadrupole moment at NLO.}}
 	\end{center}

     {\center{\bf{III. FINITE MOMENTUM TRANSFER GRAPHS AND ABOVE THRESHHOLD SCATTERING}}}

      Now it is shown how to calculate bound state graphs with finite momentum transfer.
      For pedagogical reasons we begin with a simpler graph than Fig.1. The analogous 
      one pion exchange graph shown in Fig.2 serves to illustrate $CPS$ extension to 
      the $q{\not=}0$ case.  

      Again, in the effective field theory of {\cite{1}}, we find a Born amplitude of
       \bege
       {\cal{A}}=\frac{-3eg^2M_N^3}{2f^2}{\int}{\ddp}{\ddl}
       \frac{2l_il_j-l^2{\delta}_{ij}}
	{((p-q/2)^2+{\gamma}^2)(p^2+a^2)((p+l)^2+a^2)(l^2+m_{\pi}^2)}.
	\ende
	To extract the contribution to the electric quadrupole or monopole form factor, we multiply by the 
       appropriate orthogonal tensor structure. These are $Q_{ij}=q_iq_j-\frac{q^2{\delta}_{ij}}{3}$ 
       and ${\delta}_{ij}$, respectively. Depending on which structure is desired, different integrals will
       be needed. But in both cases the master integral is  
         \begin{eqnarray}
          I &{\equiv}& {\int}{\ddp}{\ddl} \frac{1}{((p-q/2)^2+{\gamma}^2)(p^2+a^2)((p+l)^2+a^2)(l^2+m_{\pi}^2)}
	   \nonumber\\
       	    &=& \frac{1}{(4{\pi})^4}{\int}d^3xd^3y 
	      \frac{\exp{[-(a+m_{\pi})x-ay-{\gamma}|\bdra{x}-\bdra{y}|+i\frac{\bdra{q}{\cdot}(\bdra{x}-\bdra{y})} {2} ]} }
	      {x^2y|\bdra{x}-\bdra{y}|}
       \end{eqnarray}
       where the trivial delta function integrations have been performed after going to coordinate space.
       Now we write ${\bdra{q}}{\cdot}(\bdra{x}-\bdra{y}){\equiv}q|{\bdra{x}}-{\bdra{y}}|u_q$ and
        $\gt=\gamma-\frac{iqu_q}{2}$,
       where $u_q=\cos{{\theta}_q}$ will be averaged over later. In familiar fashion, the angular
       integrations are performed and a sum of 2 artificially divergent integrals remains :
       \bege
       I = \frac{1}{32{\pi}^2{\gt}}{\int}^{\infty}_0\frac{dx}{x}{\int}^{\infty}_0dye^{-(a+m_{\pi})x-ay}(e^{-{\gt}|x-y|}-e^{-{\gt}(x+y)}).
       \ende
       Using the methods developed above, and being sure to treat the two terms in an
       equivalent manner,
       the following result is quickly obtained :

	\begin{figure}
	\epsfxsize=3in
	\hbox to \hsize{\hss\epsffile{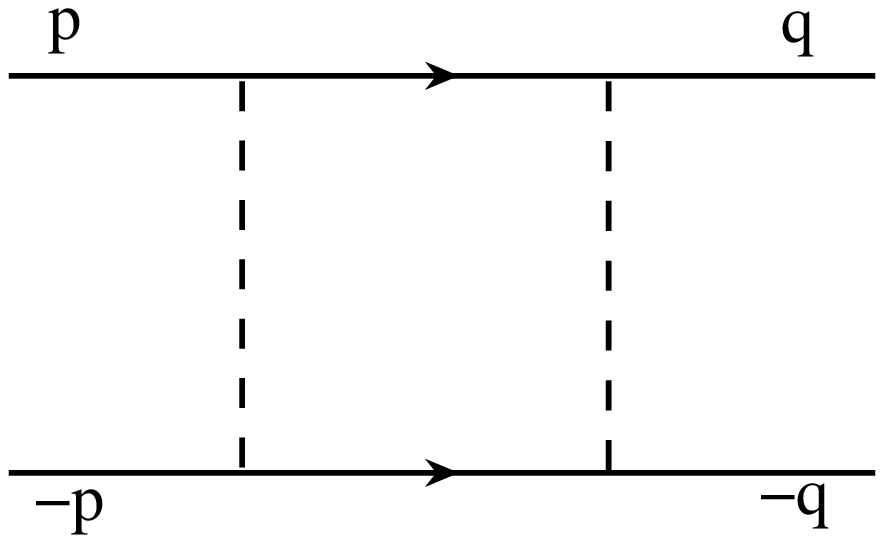}\hss}
	\label{fig:StrBrk}
	\end{figure}
	 \begin{center}
 	{\small Fig.3.{\it Box diagram contributing to the NNLO scattering amplitude.}}
 	\end{center}

       \bege
       I=\frac{1}{32{\pi}^2\gt}(\frac{1}{\gt-a}+\frac{1}{\gt+a})\log{\frac{m_{\pi}+a+\gt}{m_{\pi}+2a}}.
       \ende  
       Now we must perform the $u_q$ integration. Using eq.(23) and after some algebra the integral becomes
       \bege
       I=\frac{1}{32{\pi}^2aq}{\lbrace}{2\Im}[{\rm Polylog}_2(\frac{a-\gamma-iq/2}{m_{\pi}+2a})
       +{\rm Polylog}_2(\frac{-a-\gamma-iq/2}{m_{\pi}})]+\log{{\Bigl{[}}1+\frac{2a}{m}{\Bigl{]}}}\arctan{\frac{q}{2\gamma}}{\rbrace}
       \ende
       Although this is more complicated than the zero momentum transfer integral, eq.(8), we still obtain 
       a nice closed form solution. Similarly, we can derive an expression for the two pion exchange
       graph shown in Fig.1 at finite momentum transfer.
       The result is in terms ${\rm Polylog}_3$ and lower order polylogarithms and logarithms. 

       Now we briefly comment on above threshold scattering graphs such as the box diagram in Fig.3, where 
       $\bdra{p}$ and $\bdra{q}$ are the initial and final state momenta, and $i$ and $j$ are the spin indices for the 
       initial and final states, respectively.
        The same 
       techniques can be used for $S{\rightarrow}S$, $S{\rightarrow}D$, and $D{\rightarrow}D$ wave scattering. 
       In any case, the momenta in the numerator of the loop integral may be decoupled with the
       propagator momenta, so that only 1, 2, and 3 propagator integrals are left. The master integral is
       \bege
       I = {\int}{\ddl}\frac{1}{(l^2-p^2)((l-p)^2+m_{\pi}^2)((q-l)^2+m_{\pi}^2)}.
       \ende
       After the delta function integrations in coordinate space  
       \bege        	 
       I=\frac{1}{(4{\pi})^3}{\int}d^3xd^3y 
	      \frac{\exp{[ipy-m_{\pi}x-m_{\pi}|\bdra{x}+\bdra{y}|+i\bdra{p}{\cdot}{\bdra{x}}-i\bdra{q}{\cdot}(\bdra{x}+\bdra{y}) ]} }
	      {xy|\bdra{x}+\bdra{y}|}
	\ende
	Averaging over the angles of $p$ we have
	\bege
	I= \frac{1}{(4{\pi})^3ip}{\int}d^3xd^3y 
	      \frac{\exp{[ipy-m_{\pi}x-m_{\pi}|\bdra{x}+\bdra{y}|-i\bdra{q}{\cdot}(\bdra{x}+\bdra{y}) ]} }
	      {x^2y|\bdra{x}+\bdra{y}|}[e^{ipx}-e^{-ipx}].
	\ende
	But this is exactly the same form of eq.(31), and we can proceed almost identically. The similarity in the 
	derivation of these two loop integrals is no accident. Both for bound state graphs with finite momentum
	transferred into the loop, and for above threshhold scattering graphs, we have an external momentum 
	flow into the internal loop structure.  
	
       {\center{\bf {IV. CALCULATING ELECTROMAGNETIC PROPERTIES IN EFFECTIVE NUCLEAR FIELD THEORY }}}

       Here we only briefly review the pertinent aspects of the KSW effective field theory and the tools necessary 
       to calculate electromagetic properties. The Lagrange density with pions and nucleons is written as {\cite{1}}
       \bege
       {\cal{L}}= {\cal{L}}_0 +{\cal{L}}_1+{\cal{L}}_2+...
       \ende
       where ${\cal{L}}_n$ contains n-body nucleon operators.
       Here
       \bege
       {\cal{L}}_0 = \frac{1}{2}({\bf{E}}^2-{\bf{B}}^2) + \frac{f^2}{8}{\rm Tr}D_{\mu}{\Sigma}D^{\mu}{\Sigma}^{\dagger}
             + \frac{f^2}{4}{\lambda}{\rm Tr}m_q({\Sigma}+{\Sigma}^{\dagger}) + ...
       \ende
       where $m_q = {\rm diag}(m_u,m_d)$, $m_{\pi}^2={\lambda}(m_u+m_d)$, $f=132 {\rm  MeV}$ is the 
       pion decay constant, and the covariant derivative acting 
       on ${\Sigma}$ is 
       \bege
       D_{\mu}{\Sigma}={\partial}_{\mu}{\Sigma}+ie[Q_{em},{\Sigma}]A_{\mu}.
       \ende
       The one-body terms are 
       \bege
       {\cal{L}}_1=N^{\dagger}(iD_0+\frac{{\bf{D}}^2}{2M})N + \frac{ig_A}{2}N^{\dagger}{\bf{\sigma}}{\cdot}
             ({\xi}{\bf{D}}{\xi}^{\dagger}-{\xi}^{\dagger}{\bf{D}}{\xi})N + ... 
       \ende
       where the covariant derivative acting on the nucleon fields is 
       \bege
       D_{\mu}N = ({\partial}_{\mu}+ieQ_{em}A_{\mu})N.   
       \ende
       The two-body operators relevant to the present work is 
       \begin{eqnarray}
       {\cal{L}}_2 &=& -(C_0 + D_2{\lambda}{\rm Tr}m_q)(N^TP_iN)^{\dagger}(N^TP_iN) 
                    + \frac{C_2}{8}[(N^TP_iN)^{\dagger}(N^TP_i{\bdlra{D}^2}N) + h.c.] \nonumber\\
           &+& \frac{C_2^{S{\raw}D}}{8} {\lbrace} (N^T P_i N)^{\dagger}(N^T P_{\alpha} 
	     [{\bdlra D}_{\alpha}{\bdlra D}_i-{\bdlra D^2} \frac{{\delta}_{i{\alpha}}}{3}] N) + h.c.{\rbrace} +...
       \end{eqnarray}
       where $P_i {\equiv} \frac{1}{\sqrt{8}}{\sigma}_2{\sigma}_i{\tau}_2$ projects
       onto spin and isospin states in the spin triplet channel. Also, all of the above operators
       are assumed to be operators in the spin triplet channel, which is relevant to the 
       deuteron. Three of the counterterms have been determined previously {\cite{1}} to be 
       \bege
       C_0=-5.51{\rm fm}^2,{\rm    }   C_2=9.91{\rm fm}^4,   D_2=1.32{\rm fm}^4
       \ende 
       at renormalization scale ${\mu}=m_{\pi}$. 
       Note in eq.(43) the presence of a new local four nucleon operator which mixes $^3S_1$ and 
       $^3D_1$ wave states. This does not appear until NNLO because it scales like $1/Q$ 
       in the power counting, unlike the other two derivative coupling, $C_2$, which
       scales as $1/Q^2$. The reason is that the $C_2$ operator can be renormalized 
       by $C_0$ bubble chains on both sides whereas the $ C_2^{S{\raw}D} $ operator is only 
       renormalized on the $S$ wave side. The value of this counterterm will be determined
        in this paper. We also note that a direct quadrupole moment operator where the photon 
	couples to the four nucleon vertex does not appear until NNNLO, so we can safely neglect it.
	\begin{figure}
	\epsfxsize=4.5in
	\hbox to \hsize{\hss\epsffile{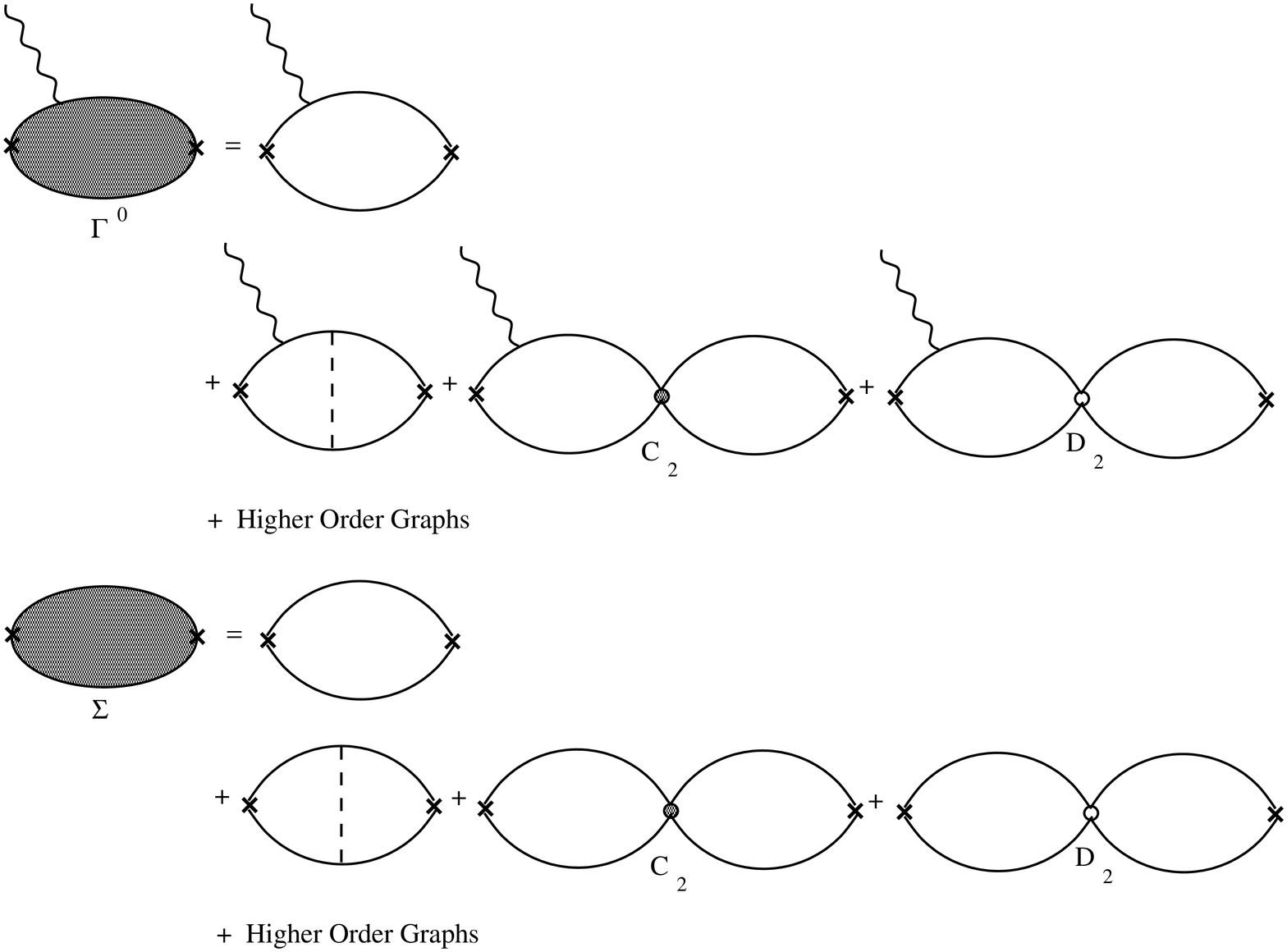}\hss}
	\label{fig:StrBrk}
	\end{figure}
	 \begin{center}
 	{\small Fig.4. {\it A graphical expansion of the irreducible three-point functions (top) and the irreducible
	two-point functions (bottom). The first line in each case is the LO amplitude while the second lines
	are the NLO amplitudes. The dashed line represent potential pions, the 'X' marks are the deuteron 
	interpolating fields which create or destroy a deuteron in the spin triplet channel.}}
 	\end{center}
	In ref.[5] the formalism necessary to calculate electromagnetic form factors in effective field
	theory were developed. 
	The matrix element of the electromagnetic current is related to the irreducible 
	2-point and 3-point functions via
	\bege
	{\la}{\bf p \prime},j|J^{\mu}_{em}|{\bf p},i{\ra} =
	        i{\Biggl{[}}\frac{{\Gamma}^{\mu}_{ij}({\bar{E}}{\bar{E}}',
		{\bf q})}{d{\Sigma}({\bar{E}})/dE}{\Biggl{]}}_{{\bar{E}},{\bar{E}}'{\raw}-B},
	\ende
         where ${\Gamma}^{\mu}_{ij} $ is the irreducible 3-point  
	 function, which is defined as the sum of all diagrams with two nucleons interacting with 
	 an external photon and each other such that the graph cannot be taken apart by cutting at a $C_0$ vertex.
	 The irreducible 2-point function ${\Sigma}$ is defined similarly, except in this case there are only two 
	 nucleons interacting with each other. The derivative of this 2-point function in eq.(45) is the 
	 wavefunction renormalization that arises from introducing the deuteron interpolating field
	 \bege
	 {\cal{D}}_i=N^T P_i N,
	 \ende                                                   
       	 which annihilates a deuteron in a definite spin and isospin state. 
	  Also note that in eq.(45) 
	 ${\bf{q}}={\bf {p'}}-{\bf{p}}$ is the photon momenta and ${\bar{E}}$ and ${\bar{E}}'$
	 are the center of mass energies.

	 The graphical expansions of the 2-point and 3-point functions up to NLO are shown in Fig.(4).
	 The zeroth component of eq.(45) is related to the definitions of the form factors by
	 \bege
	  {\la}{\bf p \prime},j|J^{0}_{em}|{\bf p},i{\ra} = e {\bigl{[}} F_C(q^2) {\delta}_{ij}+
	      \frac{1}{2M_d^2}F_Q(q^2){\bigl{(}}{\bf{q}}_i{\bf{q}}_j-\frac{1}{3}q^2{\delta}_{ij}{\bigl{)}}{\bigl{]}}.
	 \ende
	  For the present purposes, we need not consider the spatial components of this matrix element, since 
	  the quadrupole form factor $F_Q(q^2)$ is related to the quadrupole moment by 
	  \bege
	  \frac{F_Q(0)}{M_d^2}={\mu}_Q .
	  \ende

    {\center{V. THE NNLO CALCULATION OF THE DEUTERON QUADRUPOLE MOMENT}}
     
     The deuteron quadrupole moment was calculated in ref.[5] to NLO. The leading
    order contribution vanishes and the NLO result is given by
    \bege
    {\mu}_Q^{NLO} = \frac{g_A^2M(6{\gamma}^2+9{\gamma}m_{\pi}+4m_{\pi}^2)}{30{\pi}f^2(m_{\pi}+2{\gamma})^3}.
    \ende
     The numerical value for this is about $0.418 {\rm fm}^2$, which is more than ${\%}40$ larger than
     the experimental value of $0.2859 {\rm fm}^2$. This is in very rough accordance with the expectation 
     {\cite{1}} that each order in the perturbative expansion is suppressed by about ${\%}30$. A number
      of recent calculations{\cite{1,2,3,4,5,6}} suggest that the expansion does in fact converge as
       expected in calculations up to NLO. However, there has not yet been a test of the theory at NNLO.
      Thus, it would be interesting to calculate the deuteron quadrupole moment at NNLO. Since it is only 
      the second non-vanishing order, the errors are expected to be within ${\%}10$, which is rivaling the 
      accuracy of the most recent potential model calculations{\cite{9}}. Calculations using 
      potential models are consistently lower than the experimental value of the quadrupole moment 
      by ${\sim}{\%}7$. This suggests that dynamical considerations beyond potential interactions are needed 
      to accutrately describe such systems. The effective field theory approach provides the systematic 
      framework to account for these effects.
      	  
      However, as mentioned previously, there is a new direct $S{\raw}D$ wave operator at NNLO whose 
      value is unfixed. One may expect to extract this value from low energy N-N scattering data
      in the spin triplet channel. The diagrams which give rise to this ${\Delta}L=2$ $S{\raw}D$ 
      transition are shown in Fig.(5). 
            Fitting these to the triplet channel mixing parameter 
      ${\epsilon}_1$ would yield the desired value of ${C_2^{S{\raw}D}}$. However, new complications arise 
      in the NNLO above threshhold scattering channel that are not present at lower orders nor 
      in NNLO bound state problems. Work in this area is ongoing and will be presented later. 

             Alternatively, we can calculate the graphs necessary for the quadrupole moment and fit the 
       new operator to the experimental value of ${\mu}_Q$. This is the approach adopted in the present work.
       Thus, some of the questions raised in the preceding paragraph will have to wait until the 
       NNLO triplet channel scattering analysis is complete.  
       Several sources {\cite{10,11}} suggest that 
       perturbative potential pion exchange will give a large result compared 
      	\begin{figure}
	\epsfxsize=6in
	\hbox to \hsize{\hss\epsffile{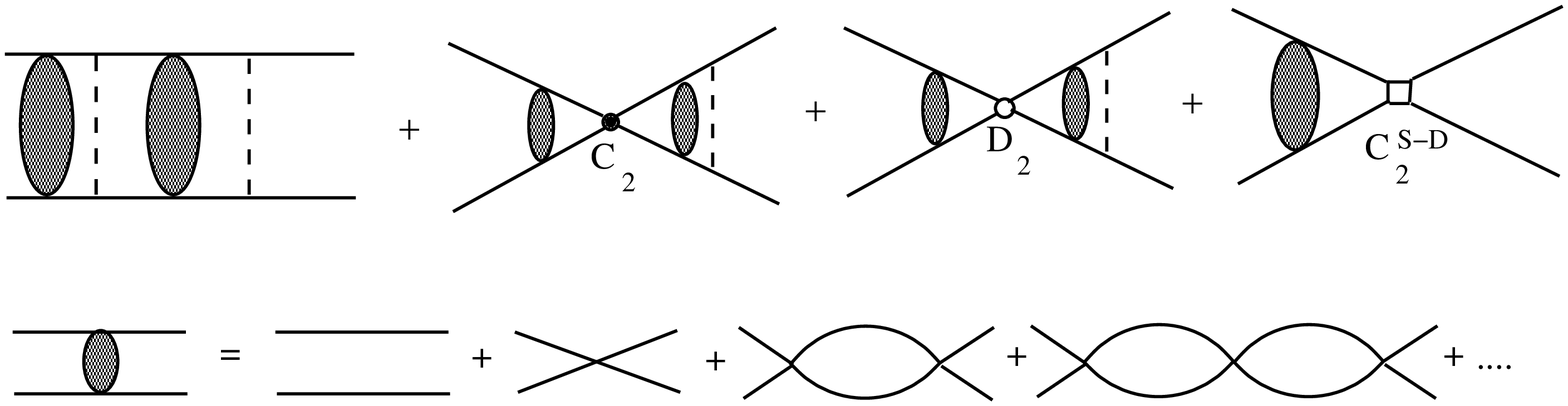}\hss}
	\label{fig:StrBrk}
	\end{figure}
	 \begin{center}
 	{\small Fig.5.{\it The top line shows the graphs contributing to the $^3S_1{\raw}^3D_1$ wave scattering
	     amplitude at NNLO. The bubble chain of $C_0$ operators is defined on the second line.}}
 	\end{center}
       with the counterterms. This 
       expectation can be verified by the results presented here.

       Expanding eq.(45) in powers of $Q$, which is the power counting scale in the KSW power counting scheme,
       we arrive at 
       \bege
       	{\la}{\bf p \prime},j|J^{0}_{em}|{\bf p},i{\ra} =
	        i{\Bigl{[}}\frac{{\Gamma}^0_{(0)}}{d{\Sigma}_{(1)}/d{\bar E}}{\Bigl{]}}
		+i{\Bigl{[}}\frac{{\Gamma}^0_{(1)}d{\Sigma}_{(1)}/d{\bar E}
		-{\Gamma}^0_{(0)}d{\Sigma}_{(2)}/d{\bar E}}{(d{\Sigma}_{(1)}/d{\bar E})^2}{\Bigl{]}}+...,
       \ende
	where we consider only the graphs that give a quadrupole
	contribution and
	\begin{eqnarray}
	{\Gamma}^0 &=& {\Gamma}^0_{(0)}+{\Gamma}^0_{(1)}+... \nonumber\\
        {\Sigma}   &=& {\Sigma}_{(1)}+{\Sigma}_{(2)}+... 
	\end{eqnarray}
	 The first term in eq.(50) is the NLO result previously calculated
	  while the other two terms are NNLO corrections.  
	A number of diagrams occur at NNLO in the effective field theory power counting. 
		  Here we are only concerned with those 
	that give rise to a quadrupole structure $Q_{ij}=q_iq_j-\frac{{\delta}_{ij}}{3}q^2$, where ${\bdra{q}}$
	is the photon 3-momenta. These diagrams must have a virtual $S{\raw}D$ and then $D{\raw}S$ transition. The graphs 
	contributing to ${\Gamma}^0_{(1)}$ are shown in Fig.(6). 
        The radiation pion refers simply to the graph evaluated by taking the pion pole in the energy integration rather than the 
	nucleon poles. While the radiation pion graph (VI) shown in Fig.(6) occurs at NNLO and has a quadrupole contribution, the 
	quadrupole piece itself is at NNNLO and can be safely neglected here. Furthermore, graph II is found to have a vanishing 
	quadrupole contribution, which can be understood as follows. Any $S{\raw}D$ transition from the first pion followed by a
	$D{\raw}S$ transition from the photon will be exactly canceled by the reverse process, namely a $S{\raw}D$ transition
	from the photon and then a $D{\raw}S$ transition via the second pion.
	In other words, the photon interacts symmetrically with the deuteron,
	whereas assymmetries in the electromagnetic interaction give rise to the quadrupole moment.
	 The remaining four graphs give non-vanishing contributions.
		
	The $C_2$ and $D_2$ graphs are directly related to the NLO one pion exchange graph in Fig.(2) by			
        \begin{figure}
	\epsfxsize=6in
	\hbox to \hsize{\hss\epsffile{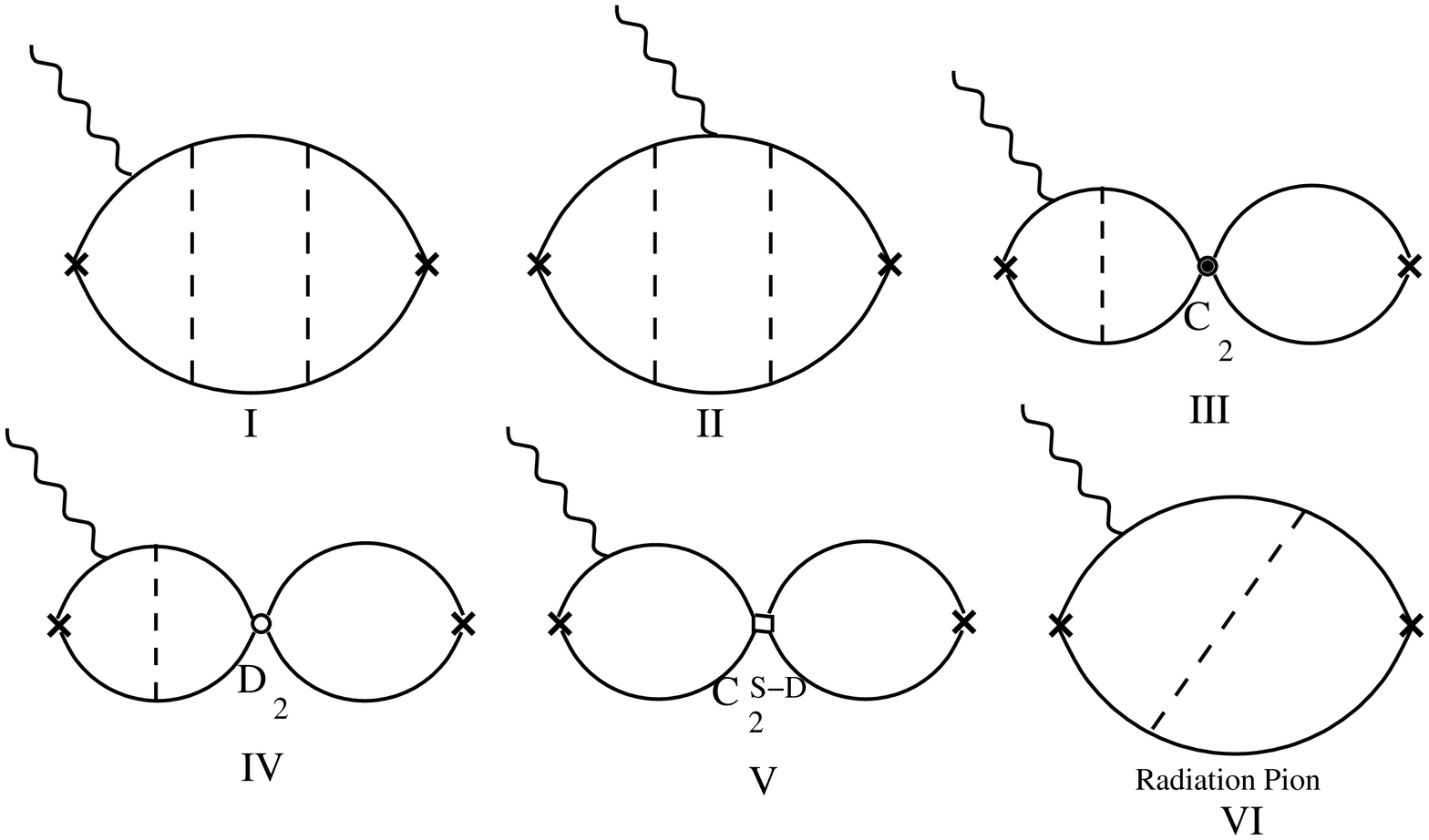}\hss}
	\label{fig:StrBrk}
	\end{figure}
	 \begin{center}
 	{\small Fig.6.{\it The graphs contributing to the NNLO calculation of the deuteron quadrupole moment. The photon
	corresponds to $A^0$.}}
 	\end{center}
	\begin{eqnarray}
	{\Gamma}^0_{III} &=& \frac{{\gamma}^2M_NC_2({\mu}-{\gamma})}{2{\pi}}{\Gamma}^0_{NLO}\nonumber\\
	{\Gamma}^0_{IV} &=& -\frac{m_{\pi}^2M_ND_2({\mu}-{\gamma})}{2{\pi}}{\Gamma}^0_{NLO}.
	\end{eqnarray}
	The $C_2^{S{\raw}D}$ graph (V) is given by 
	\bege
	{\Gamma}^0_{V} = \frac{eM_N^3C_2^{S{\raw}D}({\mu}-{\gamma})}{384{\gamma}{\pi}^2}.
	\ende
	Finally, the two potential pion exchange graph is calculated using the CPS regularization
	and renormalization scheme developed in section II and determined to be 
	\begin{eqnarray}
	{\Gamma}^0_{I} &=& -\frac{9eg^4M_N^4}{5f^4} {\Biggl{\lbrace}} {\Bigl{[}} 1280{\gamma}^9+3568{\gg}^8{\mm}+4272{\gg}^7{\mm}^2
	               +3296{\gg}^6{\mm}^3+1664{\gg}^5{\mm}^4+349{\gg}^4{\mm}^5+33{\gg}^3{\mm}^6 \nonumber\\
		&+& 60{\gg}^2{\mm}^7+18{\gg}{\mm}^8
		       -2{\mu} {\bigl{(}} 520{\gg}^8+1700{\gg}^7{\mm}+2306{\gg}^6{\mm}^2+2007{\gg}^5{\mm}^3
		       +1637{\gg}^4{\mm}^4+1215{\gg}^3{\mm}^5 \nonumber\\
		&+& 603{\gg}^2{\mm}^6+162{\gg}{\mm}^7+18{\mm}^8 {\bigl{)}} {\Bigl{]}}
		       /(294912{\pi}^3{\gg}^5({\gg}+{\mm})^2 (2{\gg}+{\mm})^3) 
		       +\frac{ {\mu}^2 ({\gg}^2+3{\mm}^2) } {24576{\pi}^3{\gg}^5} \nonumber\\
		&-&     \frac{{\mu}                \log{\frac {{\gg}+{\mm}} {2{\gg}+{\mm}} } } {3072{\pi}^3{\gg}^2}
		       +\frac{ (7{\gg}^2+{\mm}^2) \log{\frac {2{\gg}+{\mm}}       {{\mu}} } } {24576{\pi}^3{\gg}^3}
		       +\frac{{\mm}^2(3{\mm}^2-4{\gg}^2) {\bigl{(}} {\rm Polylog}_2 {\Bigl{[}} -\frac{\mm}{2{\gg}+{\mm}}
		        {\Bigl{]}} + \frac{{\pi}^2}{12} {\bigl{)}} } { 6144{\pi}^3{\gg}^5 } \nonumber\\
		&+& \frac{ (180{\gg}^5+340{\gg}^4{\mm}+409{\gg}^3{\mm}^2+12{\gg}^2{\mm}^3-213{\gg}{\mm}^4-72{\mm}^5 )
		            \log{\frac{2({\gg}+{\mm})} {2{\gg}+{\mm}} } } {73728{\pi}^3{\gg}^4(2{\gg}+{\mm})^2 } {\Biggl{\rbrace}}
        \end{eqnarray} 
	The expressions for the two-point functions are reproduced here and were calculated in {\cite{5}} to be
	\begin{eqnarray}
	\frac{d{\Sigma}_{(1)}}{d{\bar{E}}}{\Bigl{|}}_{\bar{E}=-B} &=& -i\frac{M_N^2}{8{\pi}{\gamma}} \nonumber\\
	\frac{d{\Sigma}_{(2)}}{d{\bar{E}}}{\Bigl{|}}_{\bar{E}=-B} &=& -i\frac{M_N^3}{16{\pi}^2{\gamma}}
		{\Bigl{[}}\frac{g_A^2}{2f^2}{\Bigl{(}}{\gamma}-{\mu}+\frac{m_{\pi}^2}{m_{\pi}+2{\gamma}}{\Bigl{)}}
		+D_2m_{\pi}^2({\gamma}-{\mu})-C_2{\gamma}({\mu}-{\gamma})({\mu}-2{\gamma}){\Bigl{]}}.
	\end{eqnarray}
        The partial contributions to the quadrupole moment by graphs I, III, IV, and V are
	\bege
	{\mu}_Q^I=-16.7{\rm fm}^2, {\mu}_Q^{III}=0.079{\rm fm}^2,   {\mu}_Q^{IV}=-0.096{\rm fm}^2,
	\ende
	and 
	\bege
	{\mu}_Q^V=\frac{-({\mu}-{\gamma})C_2^{S{\raw}D}M_N}{24{\pi}}.
	\ende  
	Also, the last term of eq.(50) gives an additional wavefunction renormalization contribution of 
	\begin{eqnarray}
	\tilde{\mu}_Q &=& -\frac{d{\Sigma}_{(2)}/d{\bar E}}{d{\Sigma}_{(1)}/d{\bar E}}{\mu}_Q^{NLO}\nonumber\\
		      &=& 0.203 {\rm fm}^2 .
	\end{eqnarray}
	The renormalization scale ${\mu}$ has been set to $m_{\pi}$ but it should be noted that the running 
	of the counterterms cancels the scale dependence to the order we are working. Fixing the counterterm 
	$C_2^{S{\raw}D}$ to reproduce the physical quadrupole moment yields a value of
	\bege
	 C_2^{S{\raw}D} = 5.1 {\rm fm}^4.
	\ende

	 {\center{\bf {CONCLUSIONS}}}  

     We have shown that otherwise comlicated loop diagrams arising in non-relativistic field theories, particularly 
     the EFT of Kaplan, Savage, and Wise, can be easily calculated in a new regularization and renormalization
      scheme called $CPS$. The divergences are dealt 
     with consistently by transforming them into divergences in a parameter with units of energy. This is 
     done after the angular integrations in configuration space by inserting the identity operator
     in the form ${\int}^a_{\infty}da\frac{\partial}{{\partial}a}$, where $a$ appears in the exponential. 
      For artificial divergences
     such as the artificial infrared divergences arising in Fig.1, the only requirement is consistency in dealing with 
     the infinities, since they cancel in the end. For real divergences, the $CPS$ method precisely reproduces
      the results of dimensional regularization and $PDS$ (or $\overline{MS}$) up to NLO. 
      Higher order loop integrals evaluated in $CPS$ also map onto poles in dimensional regularization. 
      This gives one the power 
      of the 
      $PDS$ subtraction scheme and the associated KSW power counting,
       without the calculational difficulties associated with dimensional regularization and
      Feynman parameters.
      As is the case with PDS, we could ignore the power law divergences and keep only 
      the four dimensional logarithmic divergences, yielding a subtraction scheme similar to $\overline{MS}$. 
      To illustrate the methods, the quadrupole moment of the deuteron is calculated to three loops,
      thus fixing the numerical value of the direct four nucleon $^3S_1{\raw}^3D_1$ operator that appears at this
      order.

      Although the emphasis in this paper has been effective field theories in nuclear physics, 
      the techniques developed are applicable in other field theories, such as non-relativistic quantum chromodynamics
      (NRQCD). 

      {\center{\bf {ACKNOWLEDGEMENTS}}}

      This work was made possible by the generous hospitality and support of the Institute for Nuclear Theory during the past
      summer. The author would like to thank Eric Swanson, Martin Savage, and Jiunn-Wei Chen for useful discussions.
       Also, I thank Gautam Rupak,
      Noam Shoresh, and Harald Griesshammer for helpful comments and suggestions.

\end{document}